# The Bean critical state: Infinitely unstable

Steven Spencer[a] and Henrik Jeldtoft Jensen[b]
Department of Mathematics, Imperial College
180 Queen's Gate, London SW7 2BZ
United Kingdom

August 24, 1995

**Abstract**

The threshold for creep in the Bean critical state is investigated. We perturb the Bean state by an energy $\Delta\epsilon$. We find that no matter how small $\Delta\epsilon$ is it will always be able to induce creep somewhere on the Bean profile. This finding has important consequences for the interpretation of low temperature creep phenomena in terms of quantum creep.

## 1 Introduction

At non-zero temperature flux lines can move due to thermal activation over energy barriers.[1, 2] The creep velocity $v$ of the flux lines will accordingly be proportional to an Arrhenius activation factor

$$v \sim \exp\{-U/k_B T\} \qquad (1)$$

where $U$ is the energy barrier, $k_B$ denotes Boltzmann's constant, and $T$ is the temperature. If all barriers $U$ are larger than some minimum barrier $U_m$ the creep will vanish exponentially for temperatures smaller than $U_m/k_B$. It is in general tacitly assumed that $U_m$ is larger than zero. Experiments which find that the creep rate does not extrapolate to zero as the temperature approaches zero have been interpreted as indicating quantum tunneling of macroscopic vortices.[3] However, if the distribution of energy barriers $D(U)$ has no gap, i.e. $U_m = 0$, detectable creep may be thermally induced at any non-zero temperature.

We demonstrate by use of computer simulations that no minimum barrier exists for two different implementations of models of the Bean state in one dimension. In the following section we introduce the models. We explain how we establish the Bean critical state and how we measure the energy barriers. We present our results and end the paper with a discussion of the implications of our model studies.

# 2  Model

Our system consists of particles interacting via a repulsive pair potential. The particles are called vortices and we denote the pair potential by $U_{vv}(r)$, where $r$ is the separation between the pair. In addition to their mutual interaction the vortices interact with a set of randomly positioned attractive pinning wells with the pinning potential denoted by $U_{vp}(r)$, where, in this case, $r$ is the distance of the vortex from the position of the pin.

We consider two different sets of interactions. The first is when the potentials take a Gaussian functional form. The vortex–vortex repulsive interaction is

$$U_{vv}^G(r) = A_v \exp(-(r/R_v)^2) \qquad (2)$$

and the vortex–pin attractive interaction is

$$U_{vp}^G(r) = -A_p \exp(-(r/R_p)^2). \qquad (3)$$

In the second, the potentials are given by truncated parabolas. Again, the vortex–vortex interaction is

$$U_{vv}^P(r) = \begin{cases} A_v(|r| - R_v)^2 & \text{if } |r| < R_v \\ 0 & \text{if } |r| \geq R_v \end{cases} \qquad (4)$$

whilst the vortex-pin interaction is

$$U_{vp}^P(r) = \begin{cases} A_p((r/R_p)^2 - 1) & \text{if } |r| < R_p \\ 0 & \text{if } |r| \geq R_p \end{cases} \qquad (5)$$

The positions $r_i$ of the vortices are restricted to the interval $[-L, L]$. The positions of the pinning centres $r_i^p$ are uniformly randomly distributed throughout the pinned region (PR) of the system, defined on the two intervals $[-L, -a]$ and $[a, L]$, leaving an unpinned region (UR) on the interval $[-a, a]$. Vortices are entered into the UR two at a time at positions $r = \pm\Delta$, with $\Delta < a$. The vortices leave the system when their positions become larger than $L$, or more precisely when $|r| > L$. The PR corresponds to the actual interior of the superconductor, whilst the mutual repulsion of the vortices in the UR, $[-a, a]$, mimics the effect of the magnetic pressure set up by the external field which makes the vortices enter a real superconductor. Working with a symmetric interval allows one to avoid the problem of defining some additional boundary force in order to push vortices into the system [4]. The number of vortices $N_v(t)$ obviously changes with time $t$, whereas the number of pinning centres $N_p$ is fixed for a specific realisation of the random potential.

At a given instant the potential energy of the system is given by

$$E_{pot} = \frac{1}{2} \sum_{i \neq j}^{N_v(t)} U_{vv}(|r_i - r_j|) + \sum_{i=1}^{N_v(t)} \sum_{j=1}^{N_p} U_{vp}(|r_i - r_j^p|). \qquad (6)$$

The model is driven in the following way. Vortices move according to the over-damped equation of motion

$$\eta \frac{dr_i}{dt} = -\frac{\partial E_{pot}}{\partial r_i}. \qquad (7)$$

Starting form a vortex free system, vortices are successively injected into the UR, $[-a, a]$. After each addition Eq. 7 is iterated until all of the vortices come to rest. We say this situation has been reached when the largest force on any one vortex is $\leq 10^{-12} A_v/R_p$. Only when all motion has ceased are new vortices injected [5].

We continue this procedure until vortices starts to leave the system at the two edges $r = \pm L$. A density gradient is set up because of the competition between the repulsive vortex–vortex interaction and the attractive vortex–pin interaction, with the vortices being most dense at the surface of the superconductor $r = \pm a$. The stability of the established vortex profile is probed by the following method.

Each of the vortices are visited successively. We displace the considered vortex, with position $r_0$, by an amount $\Delta x$, i.e. $r_0 \to r_0 + \Delta x$. We then relax the system according to Eq. 7 and then compare the vortex configurations before and after the perturbation. There are two possible outcomes:

1) The perturbation $\Delta x$ does not result in any avalanche and the perturbed vortex simply returns to its original position, $r_0$.

2) The perturbation does lead to an avalanche and a general re-arrangement of the vortex positions has occurred. We then measure the total displacement of the vortex profile given by $\delta = \sum_{n=1}^{N_v}(r_n^f - r_n^i)$, where $r_n^i$ ($r_n^f$) denotes the initial (final) position of vortex number $n$.

When we displace the vortices, we only consider perturbations in the range $0 \leq |\Delta x| \leq 1.1 R_p$ and the vortices are always displaced in the direction of decreasing vortex density. Displacements towards the high density region always result in the first outcome. For some vortices in the system, perturbations in this range only ever produce the first outcome. For the remaining vortices, a sharp threshold, $\Delta x_c$, exists. For perturbations $|\Delta x| < \Delta x_c$, outcome number 1 applies and there is no instability in the system. For $|\Delta x| \geq \Delta x_c$, the system becomes unstable and vortex re-arrangement occurs. It is also worth noting that *for all perturbations $\Delta x_c \leq |\Delta x| \leq 1.1 R_p$, the same re-arrangement, $\delta$, of the vortices occurs.* An increase in the perturbation does not give an increase in the vortex re-arrangement, $\delta$.

Our main interest is in the properties of the performed perturbation. For a vortex at position $r_0$, we calculate the threshold perturbation $\Delta x_c(r_0)$. If no instability is induced for $|\Delta x| < 1.1 R_p$ we put $\Delta x_c(r_0) = 1.1 R_p$. The increase in the energy of the system corresponding to the perturbation $\Delta x_c(r_0)$ is denoted by $\Delta \epsilon$. As we shall see in the next section there is no gap separating the measured values of $\Delta \epsilon$ from zero.

# 3   Results

We present our main results in Fig. 1. The most important point to notice is that the distribution of activation energies measured in the simulations has support all the way down to zero. The detailed shape of the distribution is model dependent, which is seen by comparing Fig. 1a and Fig. 1b. The larger support at low energies of the parabolic model is probably an effect of the sharp cut-off introduced in this model. The reason being that the forces derived from these potentials depends discontinuously upon distance. Nevertheless, both models exhibit susceptibility to perturbations of vanishing energy. It is also interesting to note that barrier distributions with seemingly diverging support at small energies have been measured experimentally on epitaxial films.[6]

The inserts in Fig. 1 show the spatial dependence of the size of the activation energies. We have plotted $\Delta\epsilon$ as function of the position $r_0$ of the vortex under consideration. One sees that the majority of avalanches are released due to perturbations at the foot, $|r_0| \approx L$, of the pile in the case of the Gaussian model. The spread in the measured values of $\Delta\epsilon$ is also much larger for vortices placed in the bottom of the pile than for vortices placed at the top. Again the model with the cut-off is different. The response of the vortex profile in this model is much more spatially homogeneous with a band of $\Delta\epsilon$ values for all positions in the pile.

Fig. 2 further illuminates the spatial dependence of the stability of the pile in the Gaussian model. The plot contains a measure of the average displacement needed in order to induce a perturbation as function of position in the pile. A word of caution is needed. As we mentioned above $\Delta x_c(r_0)$ is set equal to $1.1 R_p$ if no rearrangement in the pile was induced as a result of displacing a vortex at position $r_0$ a distance up to $1.1 R_p$. When the graph in Fig. 2 assumes the value $1.1 R_p$ this simply means that no disturbance of the profile was ever induced by displacing a particle at that position. In reality the average value of $\Delta x_c$ at such a position might of course be larger than $1.1 R_p$. The dip in $\langle \Delta x_c(r_0) \rangle$ at values of $r_0$ close to $L$ shows that the pile is most sensible to perturbations in this region.

The distribution of induced creep is shown in Fig. 3 for different levels of the activation energy. As a measure of the creep caused by a perturbation we measured (as described above) the *total* amount of displacement $\delta$ of the vortex positions. The distribution $P(\delta)$ depends not very strongly on $\Delta\epsilon$.

# 4 Discussion

Our main results of our model study are as follows:

(A) The stability of the vortex profile in the Bean state is strongly position dependent. The low density region is much more unstable than the high density region

(B) No matter how small the energy of a perturbation is it will be able to produce motion in the profile.

Since our numerical models are qualitatively similar to the physical Bean state in a real superconductor it is worthwhile exploring the relevance of our findings.

Let us first describe an experiment which will be able to assess the relevance of the spatial dependence of the stability. Apply a temperature gradient to the Bean profile in the following way. Imagine we have established the Bean state by *increasing* the external field up to a certain value. Let the temperature in the centre of the bulk of the sample be $T_b$ and the temperature at the sample surface be $T_s$. We fix the temperatures at the values $T_s = T_0$ and $T_b = T_0 + \delta$, where $\delta \ll T_0$ and $T_0$ is some appropriate low temperature. In this case we measure the creep rate $S_1$. Next we fix the temperature in the bulk at $T_b = T_0$ and the temperature at the surface at $T_s = T_0 + \delta$. We measure creep rate $S_2$. If result (A) applies we will find $S_1 \gg S_2$. This is because the most thermally active region coincides with the most unstable region when the temperature increases in the direction of decreasing vortex density.

The specific distribution of perturbation energies shown in Fig. 1 might not apply to the superconductor for several reasons. First, our models consist of particles in one dimension interacting through a short range pair potential and not three dimensional flux lines with long range interactions. Secondly, we only consider a very special subset among all possible perturbations. In reality thermal perturbations will involve the displacement of many vortices simultaneously, whereas we simply displaced a single vortex at a time. And finally, our pinning centres are relatively long ranged compared with point pinning centres in superconductors of range the superconducting coherence length. Although these differences might indeed produce a barrier distribution $D(U)$ with another functional form than the one in Fig. 1, we do not see any reason why any of the mentioned factors should introduce a gap in $D(U)$.

Let us finish with a discussion of the possible consequences of the lack of a gap in $D(U)$ for the behaviour of creep at low temperature. One can adapt a simple thermally activated flux motion (TAFM) analysis as in Ref. [7]. The creep rate $\langle S \rangle_U$ averaged over the barrier distribution is given by

$$\langle S \rangle_U = -k_B T \int_{k_B T}^{\infty} dU \frac{D(U)}{U}. \tag{8}$$

Let us assume in accordance with our finding (B) above that the behaviour at small $U$ can be described by $D(U) \sim U^{-\alpha}$ and that $D(U)$ effectively vanishes (becomes exponentially small) for $U > U_{max}$. The low temperature behaviour of the creep rate is then given by

$$\langle S \rangle_U \sim \begin{cases} -T \ln \frac{U_{max}}{k_B T} & \text{if } \alpha = 0 \\ -T^{1-\alpha} & \text{otherwise.} \end{cases} \tag{9}$$

Thus, if the barrier distribution for the superconducting Bean profile behaves like $D(U) \sim 1/U$ for small $U$ (in qualitative agreement with Fig. 1b) we find that the *thermal* creep rate approaches a constant as the temperature is lowered towards zero. This is a behaviour often observed experimentally[8, 3], though traditionally ascribed to macroscopic quantum tunneling of vortices[8, 2].

We can not claim that the behaviour of our simple one dimensional models proves that creep observed in the limit of zero temperature is thermally activated. Additional simulations in higher dimensions and of more realistic models are needed. Nevertheless, we do believe that our study shows that the perturbation spectrum of the Bean critical state is a subtle quantity. It is difficult to measure the energy barriers directly experimentally. One often has to rely on deconvolution procedures involving various assumptions concerning the nature of the time dependence of the magnetization.[8] This might explain why the barrier distributions deduced from creep experiments sometimes exhibit a lower energy cut off[8, 7] and sometimes not[6]. Experiments designed to measure the nature of the barriers in a direct way are clearly of great interest.

# 5 Acknowledgement

We are grateful to Nicola Wilkin for stimulating discussions. One of us (SS) is supported by studentship Award Ref. no. 92564742 from the EPSRC and DRA MALVERN. This work was supported by The British EPSRC under grant no. GR/36952.

a) email: s.spencer@ic.ac.uk
b) email: h.jensen@ic.ac.uk

# References

[1] M. Tinkham, *Introduction to Superconductivity* (Kriger, Malabar, Fl, 1985).

[2] G. Blatter, M.V. Feigel'man, V.B. Geshkenbein, A.I. Larkin, and V.M. Vinokur, Rev. Mod. Phys. **66**, 1125 (1994), see especially Sec. II.A.4.

[3] A number of experiments on low temperature creep is list in Sec. II.A.5 in Ref. [2].

[4] We are grateful to Oscar Pla for mentioning this very useful trick to us.

[5] A similar model has recently been studied from the view point of self-organized criticality, see R.A. Richardson, O. Pla, and F. Nori, Phys. Rev. Lett. **72**, 1268 (1894); O. Pla and F. Nori, ibid. **67**, 919 (1991).

[6] R. Griessen, C.W. Hagen, J. Lensik, and D.G. de Groot, Physica C **162-164**, 661 (1989).

[7] C.W. Hagen and R. Griessen, Phys. Rev. Lett. **62**, 2857 (1989).

[8] H.-h. Wen, H.G. Schnack, R. Griessen, B. Dam, and J. Rector, Physica C **241**, 353 (1995).

### Fig. 1a

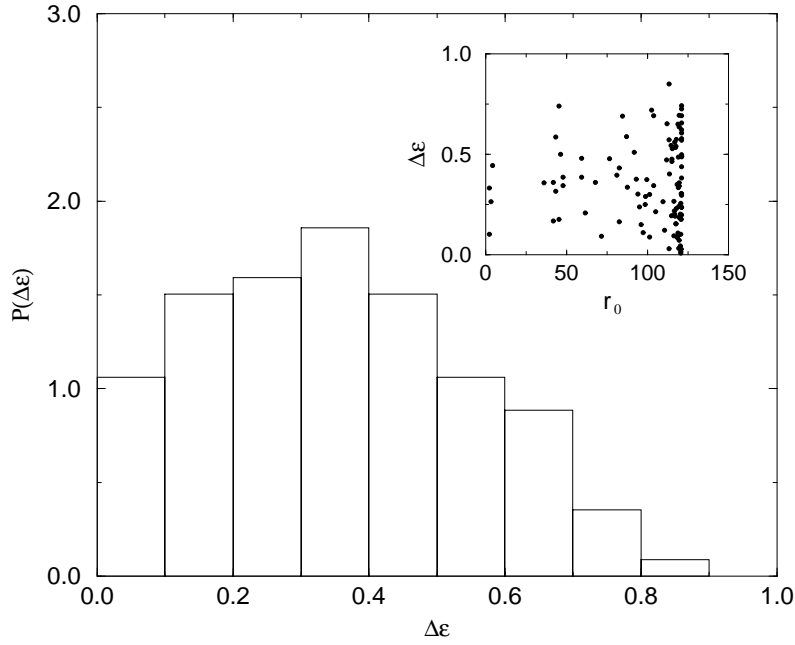

### Fig. 1b

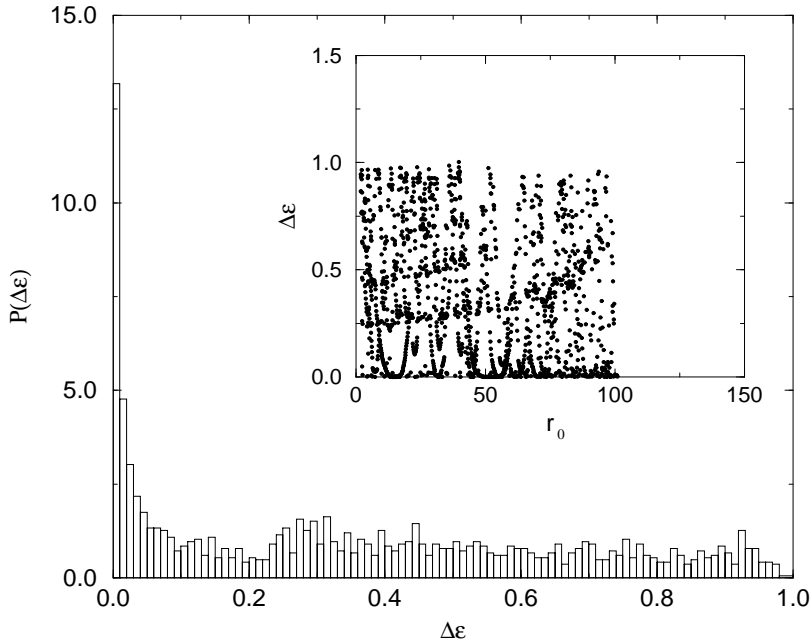

Figure 1: A) The distribution of activation energies, $P(\Delta\epsilon)$, for the Gaussian interaction model. The model parameters are $A_v = R_v = R_p = 1$ and $A_p = 0.5$. The inset shows the spatial variation of the activation energies, where $r_0 \approx 0$ is the surface of the sample and $r_0 \approx L = 120$ is in the sample bulk. B) The same as in A) except for a model with cut-off parabolic potentials. The same values of the model parameters are also used. The distribution behaves like $1/U$ for small $U$.

Fig. 2

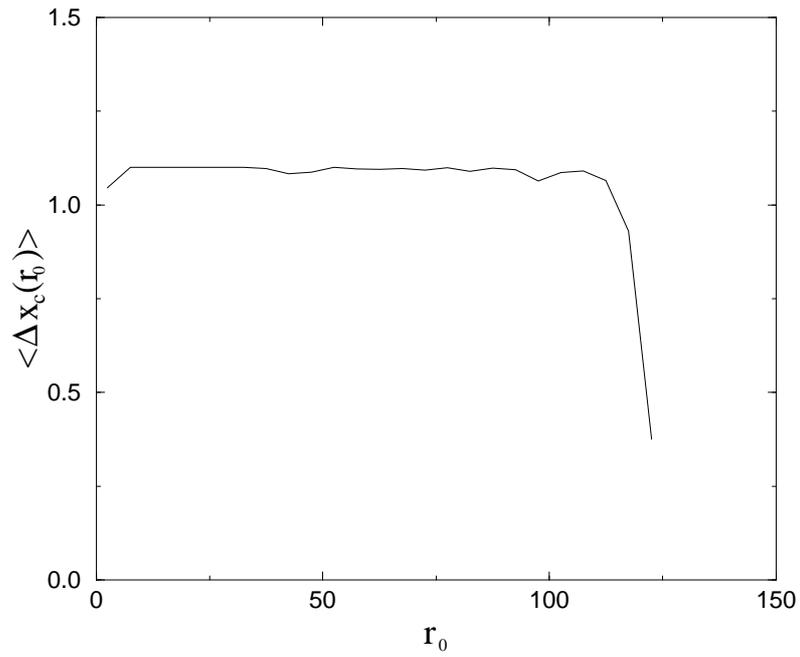

Figure 2: The average displacement of a vortex at position $r_0$ needed to cause an instability in the system, as a function of the position $r_0$. See the text for a more comprehensive explanation.

Fig. 3

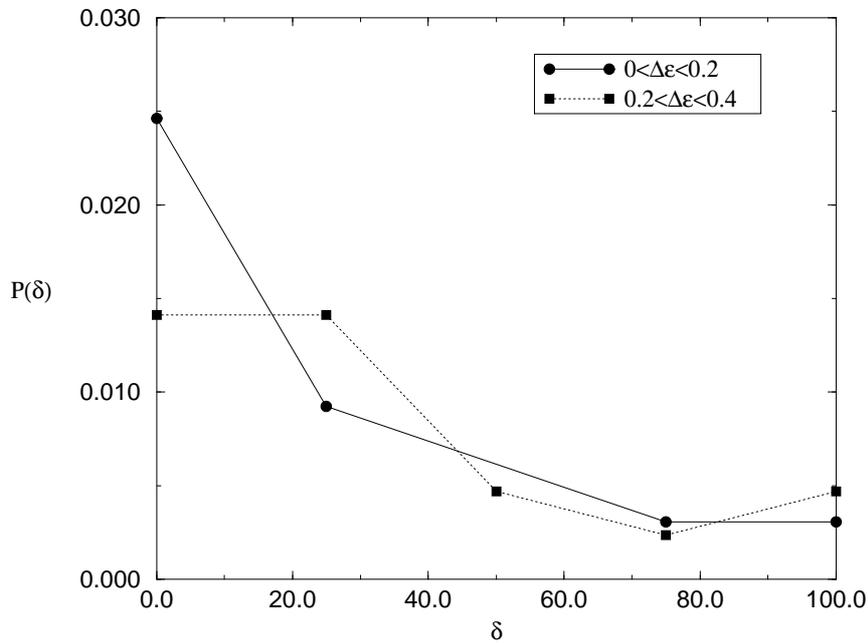

Figure 3: The distribution of the total displacement of the system after an instability has been induced. The two curves are for different ranges of the activation energies, $\Delta\epsilon$.

8